\documentclass[aps,prl,twocolumn,amsmath,amssymb]{revtex4}
\usepackage{graphicx}
\usepackage{color}
\usepackage{graphicx}
\usepackage{dcolumn}
\usepackage{bm}
\usepackage{amsmath}
\usepackage{amsfonts}
\usepackage{bbm}
\usepackage{subfigure}
\usepackage{setspace}
\newcommand{\beq}{\begin{eqnarray}}
\newcommand{\eeq}{\end{eqnarray}}

\newcommand{\bmp}{\noindent\begin{minipage}{16cm}}
\newcommand{\emp}{\end{minipage}\vskip 7mm} 

\def\drawbox#1#2{\hrule height#2pt
        \hbox{\vrule width#2pt height#1pt \kern#1pt
              \vrule width#2pt}
              \hrule height#2pt}

\def\Asym#1#2{\vcenter{\vbox{\drawbox{#1}{#2}
              \kern-#2pt 
              \drawbox{#1}{#2}}}}

\begin{document}
\title{\Large  Constraining Walking  and Custodial Technicolor}
\author{Roshan {\sc Foadi}}
\email{roshan@ifk.sdu.dk}
\author{Mads T. {\sc Frandsen}}
\email{toudal@nbi.dk}
\author{Francesco {\sc Sannino}}
\email{sannino@ifk.sdu.dk}
\affiliation{University of Southern Denmark, Campusvej 55, DK-5230 Odense M, Denmark. \\
Niels Bohr Institute, Blegdamsvej 17, DK-2100
Copenhagen, Denmark.}


\begin{abstract}
We show how to constrain the physical spectrum of walking technicolor models via precision measurements and modified Weinberg sum rules. We also study models possessing a custodial symmetry for the S parameter  at the effective Lagrangian level - custodial technicolor - and argue that these models cannot emerge from walking type dynamics. We suggest that it is possible to have a very light spin-one axial vector boson. However, in the walking  dynamics the associated vector boson is heavy while it is degenerate with the axial in custodial technicolor. 
\end{abstract}

\maketitle

Walking Technicolor (WT) is one of the major frameworks for constructing natural models able to break the electroweak symmetry dynamically \cite{Holdom:1981rm,Holdom:1984sk,Eichten:1979ah,Lane:1989ej}. It is hence important to show how to constrain phenomenologically these models using field theoretical tools, a general effective Lagrangian and the LEP data. WT makes use of near conformal dynamics. Hints of such dynamics were observed in \cite{Catterall:2007yx} for minimal walking technicolor (MWT) theories \cite{Sannino:2004qp,Dietrich:2006cm}.  The phase diagram of strongly coupled theories as function of number of colors, flavors and matter representation has been studied using the all-order (non)supersymmetric beta function in \cite{Ryttov:2007cx,Ryttov:2007sr} as well as the truncated Schwinger-Dyson Equation \cite{Dietrich:2006cm}. All the analysis point to the existence of a critical number of flavors above which asymptotically free gauge theories develop an infrared stable fixed point. 

A comprehensive low energy Lagrangian describing MWT  has been constructed in \cite{Foadi:2007ue}. It incorporates the knowledge of the underlying gauge theory via dispersion relations. More phenomenological approaches simply assume the existence of an underlying dynamics of  WT type \cite{Eichten:2007sx} and then guess the spectrum of the lightest resonances and couplings. We will show that when WT dynamics is taken into account, via the modified Weinberg sum rules (WSR) \cite{Weinberg:1967kj,Appelquist:1998xf} together with the LEP constraints \cite{Peskin:1990zt,Peskin:1991sw,Altarelli:1990zd,Altarelli:1991fk,Barbieri:2004qk} it allows us to relate the spectrum of spin-one resonances with their couplings to the weak gauge bosons. We will argue that it is not possible, within walking dynamics with a small S parameter \cite{Peskin:1990zt}, to achieve the spectrum proposed in \cite{Eichten:2007sx} and suggest what kind of strongly coupled dynamics can accommodate instead a degenerate and very light vector spectrum not at odds with precision measurements. 

We start from the observation that although WT theories are near an infrared stable fixed point, they develop the Fermi scale nonperturbatively. This implies a well defined low energy spectrum with the lightest resonances affecting directly the electroweak observables. This fact does not mean that the heavier resonances, or more generally the walking dynamics, is not relevant. Via dispersion relations the entire spectrum of the underlying theory will still affect the spectrum and couplings of the lightest ones. Our low energy spectrum consists of the lightest spin-one resonances, besides the Goldstones. The effect of walking  on the lightest spin-one resonances is modeled via modified WSRs \cite{Weinberg:1967kj,Appelquist:1998xf}. The three basic ingredients we use are: i) Asymptotic freedom of the underlying gauge theory, ii) The existence of a discrete spectrum of particles governed by the Fermi scale, iii) The effects of the walking dynamics  on the couplings and spectrum of the lowest resonances incorporated via dispersion relations.

In practice we consider a general low energy effective theory consistent with the modified sum rules and impose the associated S parameter \cite{Peskin:1990zt} to be small.  This amounts to assuming the existence of  WT with a small positive S and deduce new constraints. 
To be precise we take the value of S to be the largest possible one allowed at one sigma by experimental constraints for a heavy Higgs \cite{Barbieri:2004qk}. Minimal WT models are explicit examples possessing an intrinsic small S due to the fact that a very low number of flavors is needed to be near the conformal window \cite{Sannino:2004qp}. In \cite{Dietrich:2005jn} the reader will find the most complete catalogue of MWT and WT theories which can be used to break the electroweak symmetry with a small S.  Here we are interested in the further constraints imposed from the precision parameters proposed in \cite{Barbieri:2004qk}.  

We will be able to constrain also models proposed in \cite{Appelquist:1999dq, Duan:2000dy} which, at the effective Lagrangian level, possess an explicit {\it custodial} symmetry for the S parameter. We will refer to this class of models as custodial technicolor (CT). The new custodial symmetry is  present in the BESS models\cite{Casalbuoni:1988xm,Casalbuoni:1995yb,Casalbuoni:1995qt} which will therefore be constrained as well. In this case we expect our constraints to be similar to the ones also discussed in \cite{Casalbuoni:2007dk}.

The effective Lagrangian introduced in \cite{Foadi:2007ue} correctly describes all of the symmetries and interactions relevant for the constraints. The states present are the Goldstone bosons, their chiral partners and the lightest spin-one states. The walking dynamics is expressed by imposing the modified WSRs on the effective Lagrangian spectrum and coefficients. 
The first WSR implies:
\begin{equation}
F^2_V - F^2_A = F^2_{\pi}\ ,
\label{1rule}
\end{equation}
where $F^2_V$ and $F^2_A$ are the vector and axial mesons decay
constants.  This sum rule holds for walking and running dynamics. 
The second sum rule receives important contributions from throughout
the near conformal region and reads:
\begin{equation}
F^2_V M^2_V - F^2_A M^2_A = a\,\frac{8\pi^2}{d(R)}\,F_{\pi}^4,
\label{2rule-2}
\end{equation}
where $a$ is expected to be positive and $O(1)$ and $d(R)$ is the dimension of the representation of the underlying fermions \cite{Appelquist:1998xf,Foadi:2007ue}. In the case of running dynamics the right-hand side of the previous equation vanishes.  $a$ is a non-universal quantity depending on the details of the underlying gauge theory. $a$ is a function of the amount of walking which is the ratio of the scale above which the underlying coupling constant starts running divided by the scale below which chiral symmetry breaks. Any other approach trying to model walking should reduce to ours. We can interpolate between the walking and the running behavior of the underlying gauge theory. 

Once the hypercharge of the underlying technifermions is fixed all of the derived precision parameters defined in \cite{Barbieri:2004qk} are function solely of the gauge couplings, masses of the gauge bosons and the first excited spin-one states and one more parameter $\chi$:
\begin{eqnarray}
\hat{S} &=& \frac{(2 - \chi )\chi g^2}{2\tilde{g}^2}
      \ , \\ 
W &=& \frac{g^2}{2\tilde{g}^2}\frac{M_W^2 }{M_A^2M_V^2}{(M_A^2+(\chi-1)^2M_V^2)}
\ ,  \\
Y &=&  \frac{g'^2}{2\tilde{g}^2}\frac{M_W^2}{M_A^2M_V^2} {((1+4y^2)M_A^2+(\chi -1)^2M_V^2)}
 \ , \\
X &=& \frac{g\,g'}{2\tilde{g}^2} \, \frac{M_W^2}{M^2_A M^2_V}{ (M_A^2 - (\chi -1)^2M_V^2 )} \ .
\end{eqnarray}
$\hat{T}=\hat{U}=V=0$.  $g$ and $g^{\prime}$ are the weak and hypercharge couplings, $M_W$ the gauge boson mass, $y$ the coefficient parameterizing different hypercharge choices of the underlying technifermions \cite{Foadi:2007ue}, $\tilde{g}$ the technistrong vector mesons coupling to the Goldstones in the technicolor limit, i.e. $a=0$. It was realized in \cite{Appelquist:1999dq, Duan:2000dy} and further explored in \cite{Foadi:2007ue} that for walking theories, i.e. $a\neq 0$, the WSRs allow for a new parameter $\chi$ which in the technicolor limit reduces to $\chi_0=\tilde{g}^2 v^2/2M^2_A$ 
with $F^2_{\pi}=v^2(1-\chi^2/\chi_0)$ the electroweak vacuum expectation value and $M_{A(V)}$ the mass of the axial(vector) lightest spin-one field. To make direct contact with the WSRs and for the reader's convenience we recall the relations:\begin{eqnarray}
F^2_V =  \frac{2M^2_V}{\tilde{g}^2}\ , \quad F^2_A = 2\frac{M^2_A}{\tilde{g}^2}(1 - \chi)^2 \ .\end{eqnarray}
We have kept the leading order in the electroweak couplings over the technistrong coupling $\tilde{g}$ in the expressions above while we used the full expressions \cite{Foadi:2007ue} in making the plots.

How do we study the constraints? From the expressions above we have four independent parameters, $\tilde{g}$, $\chi$, $M_V$ and $M_A$ at the effective Lagrangian level. Imposing the first WSR and assuming a fixed value of $\hat{S}$ leaves two independent parameters which we choose to be $\tilde{g}$ and $M_A$. From the modified second WSR we read off the value of $a/d(R)$.  

\begin{figure*}[htb] 
\begin{center}
\begin{tabular}{cc}
{\includegraphics[height=5cm,width=7cm]{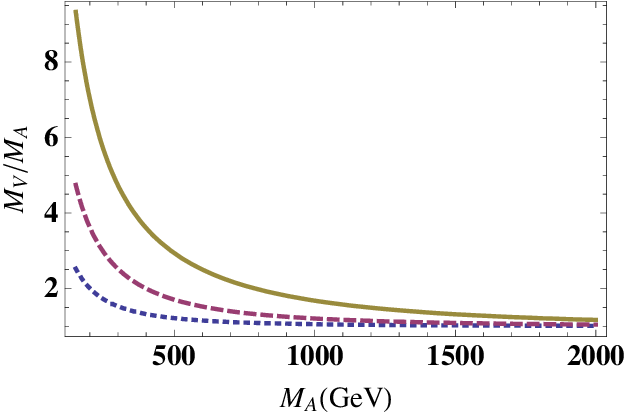}}~~~~&~~~~ 
{\includegraphics[height=5cm,width=7cm]{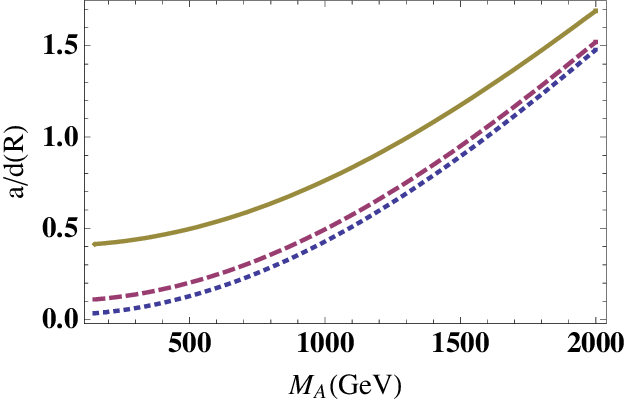}}
\end{tabular}
\end{center}
\caption{\label{fig:label2}  
In the left panel we plot the ratio of the vector over axial mass as function of the axial mass for a WT theory with an intrinsic small S parameter. The vector and axial spectrum is close only when their masses are of the order of the TeV scale and around 2 TeV and onwards the vector is lighter than the axial. The right panel shows the value $a/d(R)$ as function of the axial mass. In both plots the solid, dashed and dotted lines corresponds respectively to  $\tilde{g}=8,4,2$.
}
\end{figure*}

\vskip .2cm
{\bf Walking Technicolor}
\vskip .1cm
We will first constrain the spectrum and couplings of theories of  WT with a positive value of the $\hat{S}$ parameter compatible with the associated precision measurements at the one sigma level. More specifically we will take $\hat{S}\simeq 0.0004$ which is the highest possible value compatible with precision data for a very heavy Higgs \cite{Barbieri:2004qk}. Of course the possible presence of another sector can allow for a larger intrinsic $\hat{S}$. We are interested in the constraints coming from W and Y after having fixed $\hat{S}$.  The analysis can easily be extended to take into account sectors not included in the new strongly coupled dynamics.
\begin{figure*}[htb] 
\begin{center}
\begin{tabular}{ccc}
{\includegraphics[height=9cm,width=5cm]{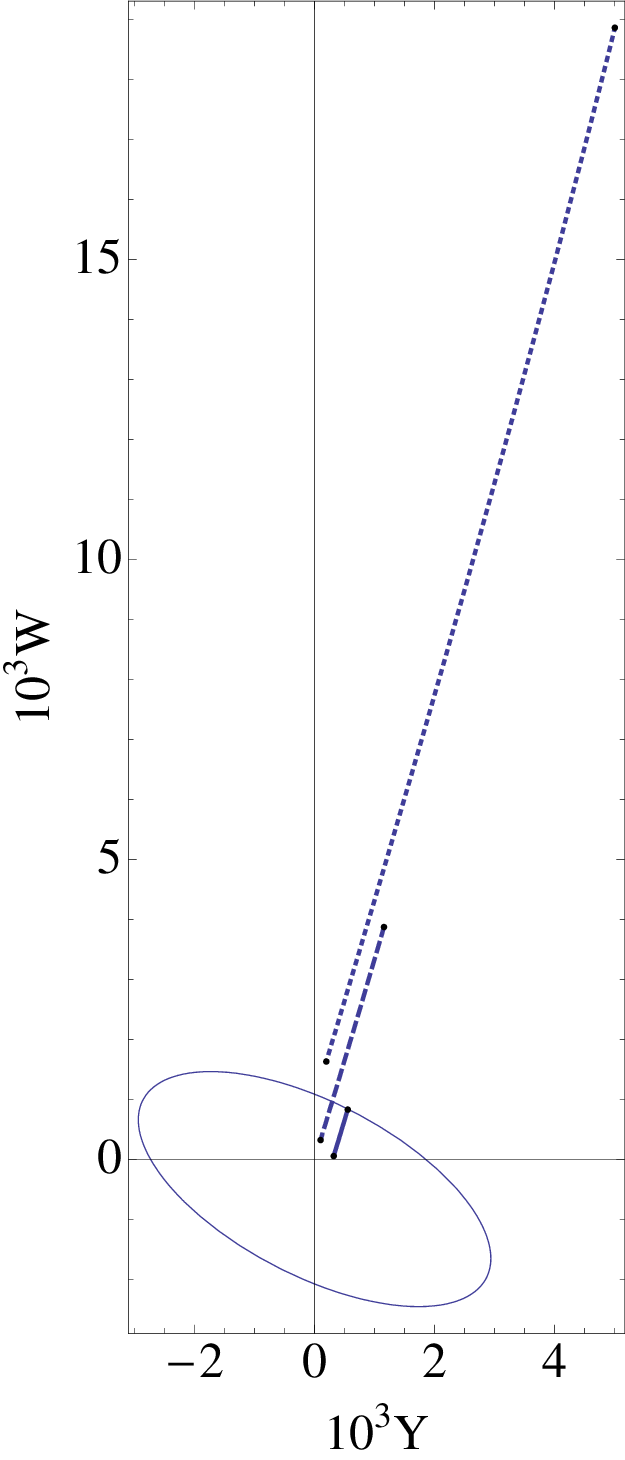}} ~~~&~~~ 
{\includegraphics[height=9cm,width=5cm]{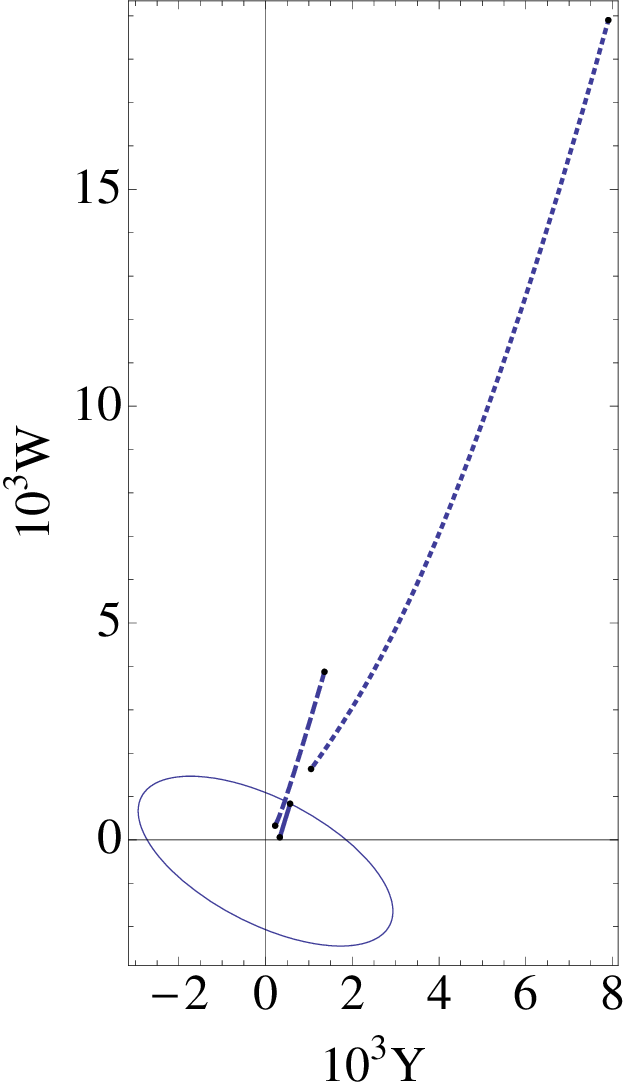}}~~~&~~~{\includegraphics[height=9cm,width=5cm]{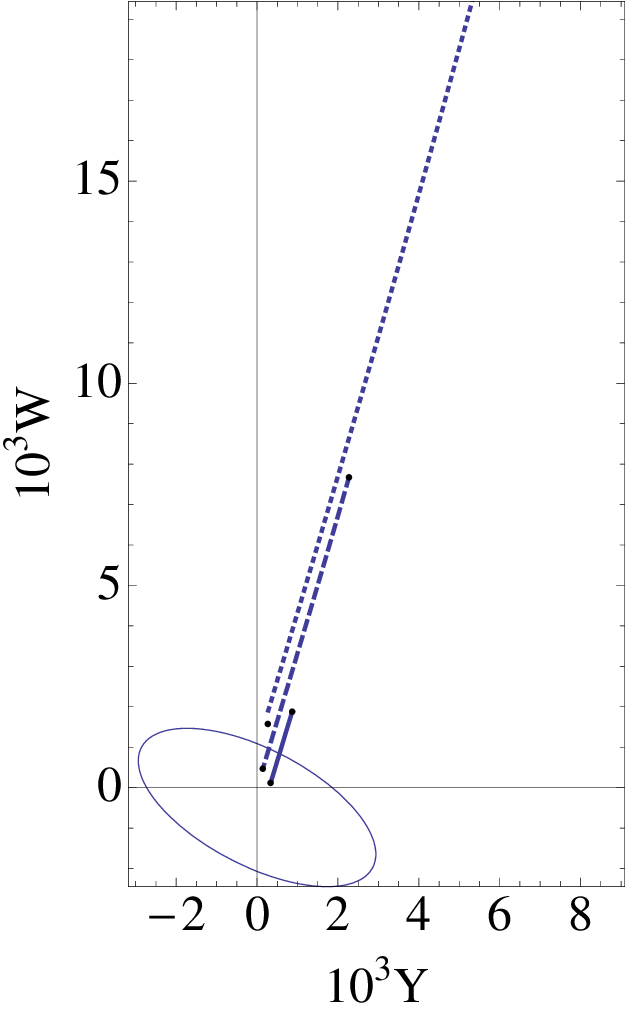}}
\\ 
~~~{WT with $y=0$}&~~~{WT with $y=1$ }&~~~CT with $y=0$
\end{tabular}
\end{center}
\caption{\label{fig:label1}  The ellipses in the WY plane corresponds to the 95\% confidence level obtained scaling the standard error ellipse axis by a 2.447 factor. The three segments, meant to be all on the top of each other,  in each plot correspond to different values of $\tilde{g}$. The  solid line corresponds to $\tilde{g}=8$, the dashed line to $\tilde{g}=4$ and the dotted one to $\tilde{g}=2$. The lines are drawn as function of $M_A$ with the point closest to the origin obtained for $M_A = 600$ GeV while the further away corresponds to $M_A=150$ GeV. We assumed $\hat{S} = 0.0004$ for WT while $\hat{S}$ is $0$ in CT by construction.  
}
\end{figure*}
The first general observation, made in \cite{Foadi:2007ue}, is that a light spin-one spectrum can be achieved only if the axial is much lighter than the associated vector meson. The second is that WT models, even with small $\hat{S}$, are sensitive to the  W-Y constraints as can be seen from the plots in Fig.~\ref{fig:label1}. Since X is a higher derivative of $\hat{S}$ it is not constraining. We find that WT dynamics with a small $\tilde{g}$ coupling and a light axial vector boson is not preferred by electroweak data. Only for values of $\tilde{g}$ larger than or about $8$ the axial vector meson can be light, i.e. of the order of $200$ GeV. However WT dynamics with a small intrinsic S parameter does not allow the spin-one vector partner to be degenerate with the light axial but  predicts it to be much heavier Fig.~\ref{fig:label2}. If the spin-one masses are very heavy then the spectrum has a standard ordering pattern, i.e. the vector meson lighter than the axial meson. We also show in Fig.~\ref{fig:label2} the associated value of $a$. We were the first to make the prediction of a very light axial vector mesons in \cite{Foadi:2007ue} on the base of the modified WSRs, even lighter than the associated vector mesons.  Eichten and Lane put forward a similar suggestion in \cite{Eichten:2007sx}.  We find that a WT dynamics alone compatible with precision electroweak data can accommodate a light spin-one axial resonance only if the associated vector partner is much heavier and in the regime of a strong $\tilde{g}$ coupling.   $a$.  We find tension with the data at a level superior to the 95\% confidence level for: a) WT models featuring $M_A \simeq M_V$ spectrum with a common and very light mass; b) WT models with an axial vector meson lighter than $300$ GeV and $\tilde{g}$ smaller than $4$, an axial vector meson with a mass lighter than or around $600$ GeV and $\tilde{g}$ smaller than $2$.  
\vskip .2cm
{\bf Custodial Technicolor}
\vskip .1cm
This is the case for which $M_A=M_V=M$ and $\chi=0$. The effective Lagrangian acquires a new symmetry, relating a vector and an axial field, which can be interpreted as a custodial symmetry for the S parameter \cite{Appelquist:1999dq, Duan:2000dy}.  The only non-zero parameters are now:
\begin{eqnarray}
W &=& \frac{g^2}{\tilde{g}^2}\frac{M_W^2 }{M^2}
\ ,  \\
Y &=&  \frac{g'^2}{2\tilde{g}^2}\frac{M_W^2}{M^2} {(2+4y^2)} \ .
\end{eqnarray}
A CT model cannot be achieved in walking dynamics and must be interpreted as a new framework. In other words CT does not respect the WSRs and hence it can only be considered as a phenomenological type model in search of a fundamental strongly coupled theory. To make our point clearer note that a degenerate spectrum of light spin-one resonances (i.e. $M<4\pi F_{\pi}$) leads to  a very large $\hat{S}=g^2 F^2_{\pi}/4M^2$. We needed only the first sum rule together with the statement of degeneracy of the spectrum to derive this $\hat{S}$ parameter. This statement is universal and it is true for WT and ordinary technicolor. 
The Eichten and Lane \cite{Eichten:2007sx} scenario of almost degenerate and very light spin-one states can only be achieved within a near CT models. A very light vector meson with a small number of techniflavors fully gauged under the electroweak can  accommodated in CT. This scenario was considered in \cite{ Zerwekh:2005wh,Zerwekh:2007pw} and our constraints apply here.

We find that in CT it is possible to have a very light and degenerate spin-one spectrum  if $\tilde{g}$ is sufficiently large, of the order say of 8 or larger as in the WT case.

We constrained the electroweak parameters intrinsic to WT or CT, however, in general other sectors may contribute to the electroweak observables, an explicit example is the new heavy lepton family introduced in \cite{Dietrich:2005jn}. 

To summarize we have suggested a way to constrain WT theories with any given S parameter. We have further constrained relevant models featuring a custodial symmetry protecting the S parameter.  When increasing the value of the S parameter while reducing the amount of walking we recover the technicolor constraints \cite{Peskin:1990zt}. We found bounds on the lightest spectrum of WT and CT theories with an intrinsically small S parameter. Our results are applicable to {\it any} dynamical model of electroweak symmetry breaking featuring near conformal dynamics \'a la walking technicolor.

\acknowledgments
We are happy to thank D.D. Dietrich, M. J\"arvinen,  K. Tuominen and T. Ryttov for discussions and/or careful reading of the manuscript. The work of R.F.,  M.T.F. and F.S. is supported by the Marie Curie Excellence Grant under contract MEXT-CT-2004-013510.

\end{document}